# Towards construction of a novel nm resolution MeV-STEM for imaging of thick biological samples

X. Yang[1*#], L. Wang[2*#], J. Maxson[3], A. Bartnik[3], M. Kaemingk[3], W. Wan[4], L. Cultrera[5], L. Wu[6], V. Smaluk[1], T. Shaftan[1], S. McSweeney[2], C. Jing[7], R. Kostin[7], Y. Zhu[6]

[1]National Synchrotron Light Source II, Brookhaven National Laboratory, Upton, NY 11973, USA
[2]Laboratory for BioMolecular Structure, Brookhaven National Laboratory, Upton, NY 11973, USA
[3]Cornell University, Ithaca, New York 14850, USA
[4]School of Physical Science and Technology, ShanghaiTech University, Shanghai 201210, China
[5]Instrumentation Division, Brookhaven National Laboratory, Upton, New York 11973, USA
[6]Condensed Matter Physics and Materials Science Division, Brookhaven National Laboratory, Upton, NY 11973, USA
[7]Euclid Techlabs LLC, 365 Remington Blvd., Bolingbrook, IL, USA
* These authors contributed equally.
# Corresponding authors: xiyang@bnl.gov, lwang1@bnl.gov

## ABSTRACT

Driven by life-science applications, mega-electron-volt Scanning Transmission Electron Microscope (MeV-STEM) has been proposed here to image thick biological samples as conventional Transmission Electron Microscope (TEM) may not be suitable to image samples thicker than 300-500 nm and various volume electron microscopy (EM) techniques either suffering from low resolution, or low speed. The high penetration of inelastic scattering signals of MeV electrons could make the MeV-STEM an appropriate microscope for biological samples as thick as 10 µm or more with a nanoscale resolution, considering the effect of electron energy, beam broadening and low-dose limit on resolution. The best resolution is inversely related to the sample thickness and changes from 6 nm to 24 nm when the sample thickness increases from 1 µm to 10 µm. To achieve such a resolution in STEM, the imaging electrons must be focused on the specimen with a nm size and a mrad semi-convergence angle. This requires an electron beam emittance of a few picometer, which is ~1,000 times smaller than the presently achieved nm emittance, in conjunction with less than $10^{-4}$ energy spread and 1 nA current. We numerically simulated two different approaches that are potentially applicable to build a compact MeV-STEM instrument: 1) DC gun, aperture, Superconducting radio frequency (SRF) cavities, and STEM column; 2) SRF gun, aperture, SRF cavities, and STEM column. Beam dynamic simulations show promising results, which meet the needs of an MeV-STEM, a few picometer emittance, less than $10^{-4}$ energy spread, and 0.1-1 nA current from both options. Also, we designed a compact STEM column based on permanent quadrupole quintuplet not only to demagnify the beam size from 1 µm at the source point to 2 nm at the specimen, but also to provide the freedom of changing the magnifications at the specimen and a scanning system to raster the electron beam across the sample with a step size of 2 nm and the repetition rate of 1 MHz. This makes it possible to build a compact MeV-STEM and use it to study thick, large-volume samples in cell biology.

# INTRODUCTION

Predictive power on living organisms requires a deep knowledge of cellular structure, which includes both the hierarchical organization and the interaction of components, toward a full understanding of biological function across different length scales. These insights are essential for predicting and controlling biological function in support of the Department of Energy (DOE) research missions and life-science applications. One appealing method to provide such knowledge with nanometer resolution is cryo-Electron Tomography (cryo-ET), where biological samples are fast frozen to preserve their native hydrated states, then, imaged at liquid nitrogen temperature to reduce radiation damage during the measurement. Cryo-ET has been successfully employed to resolve subcellular structures and study dynamic processes inside cells and tissues, with the capability of achieving nanometer resolution for the entire sample as well as atomic resolution for abundant molecular components *in-situ* through averaging those structures [1-7].

Due to the low electron dose required by biological samples to reduce the radiation damage and the short electron elastic mean free path at 100-300 keV, cryo-ET limits the sample thickness to 300-500 nm [8-11]. To visualize large/thick biological samples (e.g., cells and tissues), various volume electron microscopy (EM) techniques have been developed [12,13]. In Array Tomography, sections prepared by microtome, cryo-microtome or ultramicrotome are imaged using scanning electron microscopes (SEMs) or TEMs. The resolution is limited by the section thickness ($\geq$ 60 nm). Alternatively, the top surface is imaged after cutting with an ultramicrotome in serial block face SEM (SBF-SEM) or with a focused ion beam (FIB) in FIB-SEM. In both SBF-SEM and FIB-SEM, the resolution is limited by the slice thickness ($\sim 30 - 100$ nm).

To ensure that biological samples are as close as possible to their native state, cryo-FIB-SEM has been developed [14-18]. Resolution in depth can be as high as a few nanometers. However, depending on the imaging parameters, the imaging process often takes 15-25 seconds and the milling process takes 7-36 seconds [15]. A total of 15-20 hours is required to image a 10 μm thick sample with a slice thickness of 8 nm and an area of $3 \times 2$ μm$^2$ (8 nm pixel size) [18]. In addition, there are charging artifacts due to positively charged lipid deposits, curtaining artifacts due to density and content changes, and linear artifacts due to the milling process [14,16]. Thus, being able to study thick samples rapidly and efficiently while maintaining nanoscale resolution is highly desirable, as it will significantly increase the rate of scientific discoveries.

In cryo-EM or cryo-ET, with the assumption of 'Weak Phase Object Approximation' (WPOA), the one-to-one mapping between the exit electron wave function and the projected potential of the specimen can be established. The image is mainly formed by coherent interference (i.e., elastic scattering) among the electron waves (i.e., the phase contrast, which strongly depends on defocus, spherical aberration, objective-aperture size, and illumination conditions). The elastically scattered electrons contribute to the signal while inelastically scattered electrons contribute to the background noise, thus decreasing the signal-to-noise ratio (SNR). As shown by M. Du *et al* [19], when the ice is thicker than 1.7 μm in a TEM, there are essentially no elastically scattered electrons within the detectable angular range at 300 keV. Thus, conventional cryo-EM and cryo-ET are not suitable to study thick samples using WPOA.

As shown [20-23], STEM has been employed to study biological samples up to 1,000 nm thick, as opposed to 300 nm thick biological samples imaged by cryo-EM and cryo-ET using 200-300 keV TEMs [22]. In STEM, the electron beam is focused on the plane of the specimen, and there is no image-magnifying lens behind the specimen. The image is formed by mapping detector counts point-by-point to scan positions. The bright-field (BF), annular dark-field (ADF), and high-angle annular DF (HAADF) images are formed by collecting electrons within small (e.g. $< 10$ mrad), medium (e.g. $10 - 50$ mrad), and high (e.g. $> 50$ mrad) angles, respectively [24-26]. For cryo-samples, objects with higher concentrations of heavy atoms will be preferentially detected by ADF imaging, forming the so-called "Z-contrast" (Z is the atomic number). The mass density variations in different parts of the sample are detected by BF imaging.

As the allowable sample thickness depends on the electron energy and image formation mechanism, we propose to implement a mega-electron-volt Scanning Transmission Electron microscope (MeV-STEM) based on amplitude contrast instead of phase contrast. Comparing a STEM with a TEM, the main difference lies in how the resulting image is formed. In a TEM for structural biology, an objective lens focuses the transmitted electrons to form an image based on phase contrast. In a STEM, a focused electron beam scans across the sample in a raster pattern, and the transmitted electrons are collected to form an amplitude contrast image. In a STEM, any severe energy loss and large angle deviation occurred during electron thick-sample interaction won't affect the detected signals since there exists no lens between the sample and the detector; however, in a TEM, those effects would have a significant impact on the image resolution via chromatic and spherical aberrations. Thus, STEM is not limited to the sample to be ultra-thin, enabling the study of thicker samples or 3D materials. However, a full understanding of electron scattering and attainable spatial resolution in biological samples that are a-few-micron thick or more represents a highly intricate undertaking, necessitating substantial dedication and resources. It is important to note that delving into these intricacies is beyond the scope of this paper.

As we will discuss in the STEM imaging below, at 300 keV, the unscattered, single elastic scattered and multiple (named plural) elastic scattered electrons emerging from the sample drop to less than 1% while the inelastic scattered electrons stay at 31% after a 1-μm thick ice layer. Furthermore, the inelastic scattered electrons are still above 1% after passing through a 4-μm thick ice layer. If the electron energy is increased to 3 MeV, the inelastically scattered electrons will remain at 63% after passing through a 10-μm thick ice layer and 40% after passing through a 20-μm thick ice layer. Thus, compared to conventional TEM, STEM can image much thicker samples ($\geq 10$ μm). This makes MeV-STEM extremely appealing to life-science applications: MeV-STEM could be used to study intact cells with sample thickness 10 μm or more. For the first time, such a system will be able to rapidly and efficiently image large biological samples (subcellular structures, biomolecular components and bio-reactions in cells) while maintaining nanometer resolution ($6 - 24$ nm).

With conventional round lenses, the 3 MeV electron microscope needs to be more than 10 m tall, and the lens has usually 1 m in diameter [27]. Instead, we designed a compact STEM column based on the novel quadrupole quintuplets pioneered at Brookhaven National Laboratory for MeV electron microscopes [28]. With the permanent quadrupole quintuplet, the lens diameter is only around a centimeter, and the microscope can fit in a small laboratory. Such MeV-STEM column will not only demagnify the beam size but also have the freedom of changing the beam size at the specimen, together with an adjustable aperture, in a broad range from 2 nm to 16 μm.

## RESULTS
## I. DESIGN PARAMETERS OF ACHIEVING 6-24 nm RESOLUTION FOR 1-10 m THICK BIO-SAMPLE
### 1.1 Electron cross sections

The angular distribution of scattering from a target atom can be described by the differential scattering cross-section. For elastic scattering, the differential cross-section follows the Rutherford formula for the screened Coulomb potential of the nuclear charge. In the Wentzel approximation, the differential cross-section for elastic scattering in first order Born approximation becomes [22]:

$$\frac{d\sigma_{el}}{d\Omega} = \left[\frac{2ZR^2\left(1+\frac{E}{E_0}\right)}{a_H(1+(\frac{\theta}{\theta_0})^2)}\right]^2 , \quad \theta_0 = \frac{\lambda}{2\pi R} , \quad R = a_H Z^{-1/3} \tag{1}$$

where $\sigma_{el}$ is the elastic scattering cross-section, $\Omega$ is the solid angle, Z is the atomic number, $E$ is the electron energy, $E_0$ is the rest energy of the electron, $a_H$ is the Bohr radius (0.0529 nm), $\theta$ is the scattering angle, $\theta_0$ is the characteristic scattering angle below which 50% of the electrons are scattered into, $\lambda$ is the electron wavelength. The characteristic scattering angle for oxygen is 15 mrad for 200 keV electrons. Integrating Eq. 1 yields the total cross-section [22]:

$$\sigma_{el} \approx \frac{h^2 c^2 Z^{4/3}}{\pi E_0^2 \beta^2} , \quad \beta^2 = 1 - \left[\frac{E_0}{E+E_0}\right]^2 \tag{2}$$

The angular dependence and the cross-section of inelastic scattering can be approximated with a Bethe-model [22,29]:

$$\frac{d\sigma_{inel}}{d\Omega} \approx \frac{Z\lambda^4\left(1+\frac{E}{E_0}\right)^2}{4\pi^2 a_H^2} \left[\frac{1-\left(1+\frac{\theta^2}{\theta_0^2}\right)^{-2}}{(\theta+\theta_E^2)^2}\right] , \quad \theta_E = \frac{\Delta E}{E}\frac{E+E_0}{E+2E_0} \tag{3}$$

where $\sigma_{inel}$ is the inelastic scattering cross-section, $\theta_E$ is the characteristic angle that is responsible for the decay of the inelastic scattering, $\Delta E$ is the mean energy loss from a single inelastic scattering event (e.g., 39.3 eV for amorphous ice [19]). An inelastic scattering is concentrated within much smaller angles than elastic scattering. The characteristic angle $\theta_E$ is typically of the order of 0.1 mrad for 200 keV electrons.

The total cross section for scattering from amorphous ice is [30]

$$\sigma_{ice} = \sigma_{O\_atom} + 2\sigma_{H\_atom} \tag{4}$$

where $\sigma_{O\_atom}$ and $\sigma_{H\_atom}$ are the scattering cross sections of Oxygen and Hydrogen atoms in amorphous ice. As pointed out by Jacobsen *et al* [31], the inelastic scattering of hydrogen was not accurate. Thus, the approximation proposed by Jacobsen *et al* was followed.

## 1.2 Electron interaction probabilities

The chance of an electron undergoing event $i$ (e.g., elastic scattering, inelastic scattering) within a sample thickness $dt$ is given by

$$P = \sigma_i \rho dt = K_i dt, \tag{5}$$

where $\sigma_i$ is the cross section for event $i$, and $\rho$ is the sample density, $K_i$ is the scattering coefficient.

After interacting with the sample, the electron would fall into the following five categories [19,31] (see Fig. 1a), depending on the respective scattering coefficients (Eq. 7):

$I_{noscat}$ designates electrons undergoing no scattering;
$I_{1el}$ designates electrons being elastically scattered once, remaining within the detector collection angles;
$I_{el,plural}$ designates electrons undergoing multiple elastic scatterings without any inelastic scatterings and remaining within the detector collection angles;
$I_{inel}$ designates electrons undergoing at least one inelastic scattering and remaining within the detector collection angles;
$I_{out}$ designates electrons being scattered outside the detector collection angles.

The sum of these five types of electrons is equal to the incident electron beam intensity $I_0$:

$$I_{noscat} + I_{1el} + I_{el,plural} + I_{inel} + I_{out} = I_0 \tag{6}$$

and the corresponding scattering coefficients are:

$$K_{el} = K_{el,in} + K_{el,out}$$

$$K_{inel} = K_{inel,in} + K_{inel,out}$$

$$K_{out} = K_{el,out} + K_{inel,out} \tag{7}$$

where $K_{I,in}$ is the scattering coefficients for category $I$ ('$el$' including single to multiple elastic scattering and '$inel$' include at least one inelastic scattering) within the detector collection angles and can be estimated by numerically integrating Equations 1 and 3 then multiplying the sample density (Eq. 5), and $K_I$ can be estimated by numerically integrating Equations 1 and 3 over all possible angles. The cross sections of Oxygen in amorphous ice are calculated using Equations 1-3 and listed in Table 1. For detector collection angles from 0 to 10 mrad, the scattering coefficients for Oxygen in amorphous ice are calculated using Eq. 7 and results listed in Table 2.

| | | Elastic cross section (nm²) | | | | | Inelastic cross section (nm²) | | | |
|---|---|---|---|---|---|---|---|---|---|---|
| | | Detector collection angle | | | | | Detector collection angle | | | |
| Electron energy (eV) | $\theta_0$ (mrad) | 0-10 mrad | 10-50 mrad | 50-100 mrad | Total | $\theta_E$ (mrad) | 0-10 mrad | 10-50 mrad | 50-100 mrad | Total |

| | | | | | | | | | |
|---|---|---|---|---|---|---|---|---|---|
| 100,000 | 22.3 | 1.7E-05 | 6.7E-05 | 1.6E-05 | 1.0E-04 | 0.214 | 1.6E-04 | 3.1E-05 | 2.4E-06 | 1.9E-04 |
| 200,000 | 15.1 | 1.9E-05 | 3.8E-05 | 5.1E-06 | 6.2E-05 | 0.114 | 1.1E-04 | 1.2E-05 | 6.9E-07 | 1.3E-04 |
| 300,000 | 11.8 | 2.1E-05 | 2.6E-05 | 2.6E-06 | 5.0E-05 | 0.080 | 9.4E-05 | 6.8E-06 | 3.4E-07 | 1.0E-04 |
| 1,000,000 | 5.2 | 2.7E-05 | 6.8E-06 | 3.6E-07 | 3.4E-05 | 0.029 | 5.9E-05 | 1.1E-06 | 4.6E-08 | 6.0E-05 |
| 3,000,000 | 2.1 | 2.9E-05 | 1.3E-06 | 5.5E-08 | 3.1E-05 | 0.011 | 4.1E-05 | 1.7E-07 | 6.9E-09 | 4.1E-05 |

Table 1. Electron scattering cross section for Oxygen from 100 keV to 3 MeV. $\theta_0$ and $\theta_E$ are the characteristic scattering angles for elastic and inelastic scattering.

| Electron energy (eV) | $K_{el,in}$ | $K_{el,out}$ | $K_{el}$ | $K_{inel,in}$ | $K_{inel,out}$ | $K_{inel}$ | $K_{out}$ |
|---|---|---|---|---|---|---|---|
| 100,000 | 5.23E-04 | 2.55E-03 | 3.08E-03 | 4.95E-03 | 1.04E-03 | 5.99E-03 | 3.59E-03 |
| 200,000 | 5.90E-04 | 1.32E-03 | 1.91E-03 | 3.51E-03 | 3.95E-04 | 3.91E-03 | 1.71E-03 |
| 300,000 | 6.44E-04 | 8.87E-04 | 1.53E-03 | 2.91E-03 | 2.20E-04 | 3.13E-03 | 1.11E-03 |
| 1,000,000 | 8.20E-04 | 2.21E-04 | 1.04E-03 | 1.81E-03 | 3.45E-05 | 1.85E-03 | 2.56E-04 |
| 3,000,000 | 9.00E-04 | 4.08E-05 | 9.41E-04 | 1.26E-03 | 5.32E-06 | 1.27E-03 | 4.61E-05 |

Table 2. Scattering coefficients for electrons to be scattered from 100 keV to 3 MeV with a collection angle from 0 to 10 mrad (nm$^{-1}$).

### 1.3 Electron intensity change while traveling through the specimen

The intensity change of electrons in each category after passing through a sample slice of thickness *dt* is

(1) Unscattered:
$$dI_{noscat} = -I_{noscat}(K_{inel} + K_{el})dt \qquad (8)$$
where the initial condition is $I_{noscat}(0) = I_0$

(2) Detected single elastic scattered electrons:
$$dI_{1el} = I_{noscat}K_{el,in}dt - I_{1el}(K_{inel} + K_{el})dt \qquad (9)$$

(3) Detected multiple elastically scattered electrons:
$$dI_{el,plural} = I_{1el}K_{el,in}dt - I_{el,plural}(K_{out} + K_{inel,in})dt \qquad (10)$$

(4) Detected inelastic scattered electrons:
$$dI_{inel} = (I_{noscat} + I_{1el} + I_{el,plural})K_{inel,in}dt - I_{inel}(K_{out})dt \qquad (11)$$

(5) Undetected electrons:
$$dI_{out} = (I_0 - I_{out})K_{out}dt, \qquad (12)$$
where the initial condition is $I_{out}(0) = 0$.

Although these expressions possess limitations in capturing certain nuances of plural scattering (e.g., the probability that scattered-out electrons might be scattered back and fall within the detector collection angles, the direction change of scattered electrons after passing through the sample), they serve as a valuable initial approximation. A plot of the electron distributions in the five categories is shown in Fig. 1. As shown in Fig. 1b, the single elastic scattered electrons (blue dashed line) and plural (multiple) elastic scattered electrons (black dashed line) emerging from the sample drop to less than 1% after a 1-μm thick ice layer, and the inelastic scattered electrons (red dashed line) drop below 1% after a 4-μm thick ice layer at 300 keV. However, if the electron energy increases to 3 MeV, the single elastic scattered electrons (blue solid line) drop below 1% after a 2-μm thick ice layer (sample thickness doubled), while the plural scattered electrons (black solid line) drop below 1% after a 3.5-μm thick ice layer. The inelastically scattered electrons still stay at 63% after a 10-μm thick ice layer and 40% after a 20-μm thick ice layer, shown as the red solid line curve. In addition, Fig. 1d clearly indicates the benefit of increasing the electron beam energy – the characteristic angle of elastic scattering ($\theta_0$) decreases from 22.3 mrad to 2.1 mrad (a factor of 10) with the increase of beam energy from 100 keV to 3 MeV; hence, a detector with fixed angles (e.g., 0-10 mrad) collects much more multiple scattering events as signals in the 3 MeV case compared to low energy cases. Such difference increases with the increase of sample thickness (see Fig. 1c). Therefore, an MeV-STEM is capable of imaging much thicker samples (10 μm or thicker) than a conventional TEM, with a compromised resolution depending on the sample thickness.

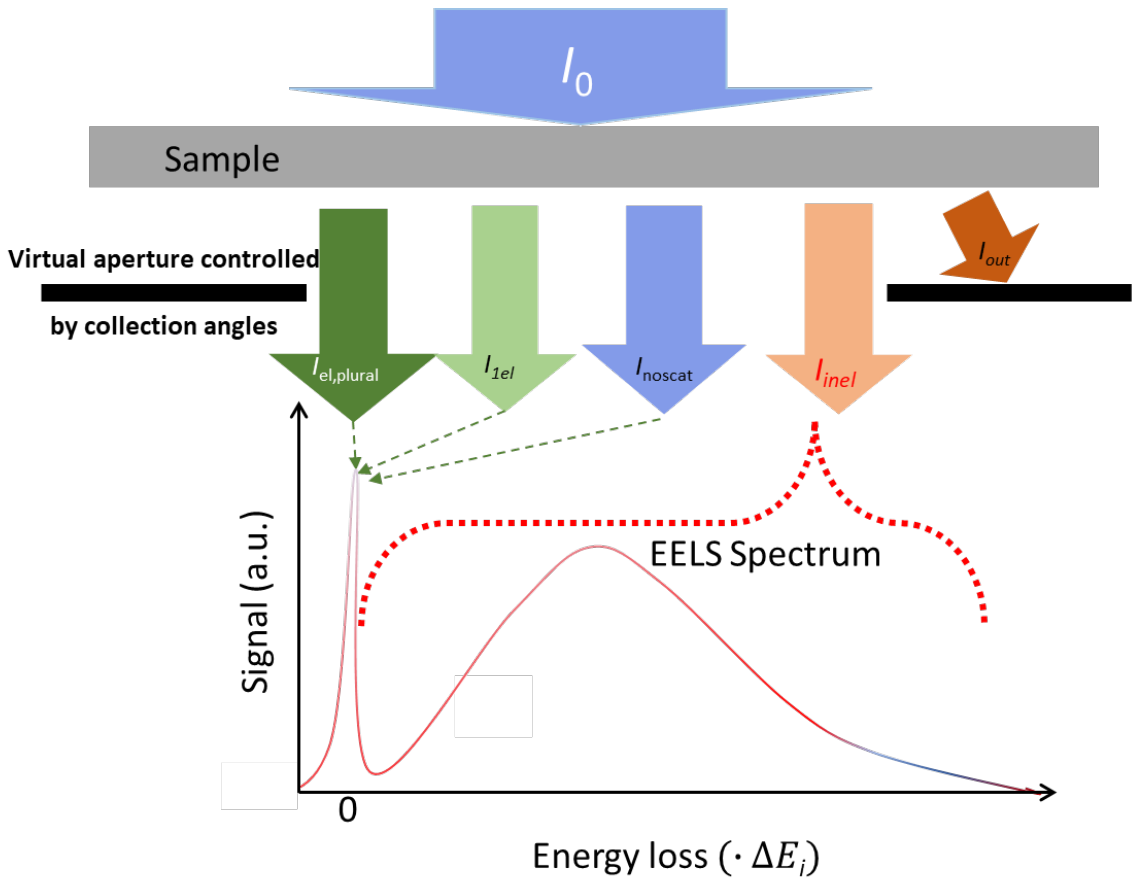

(a)

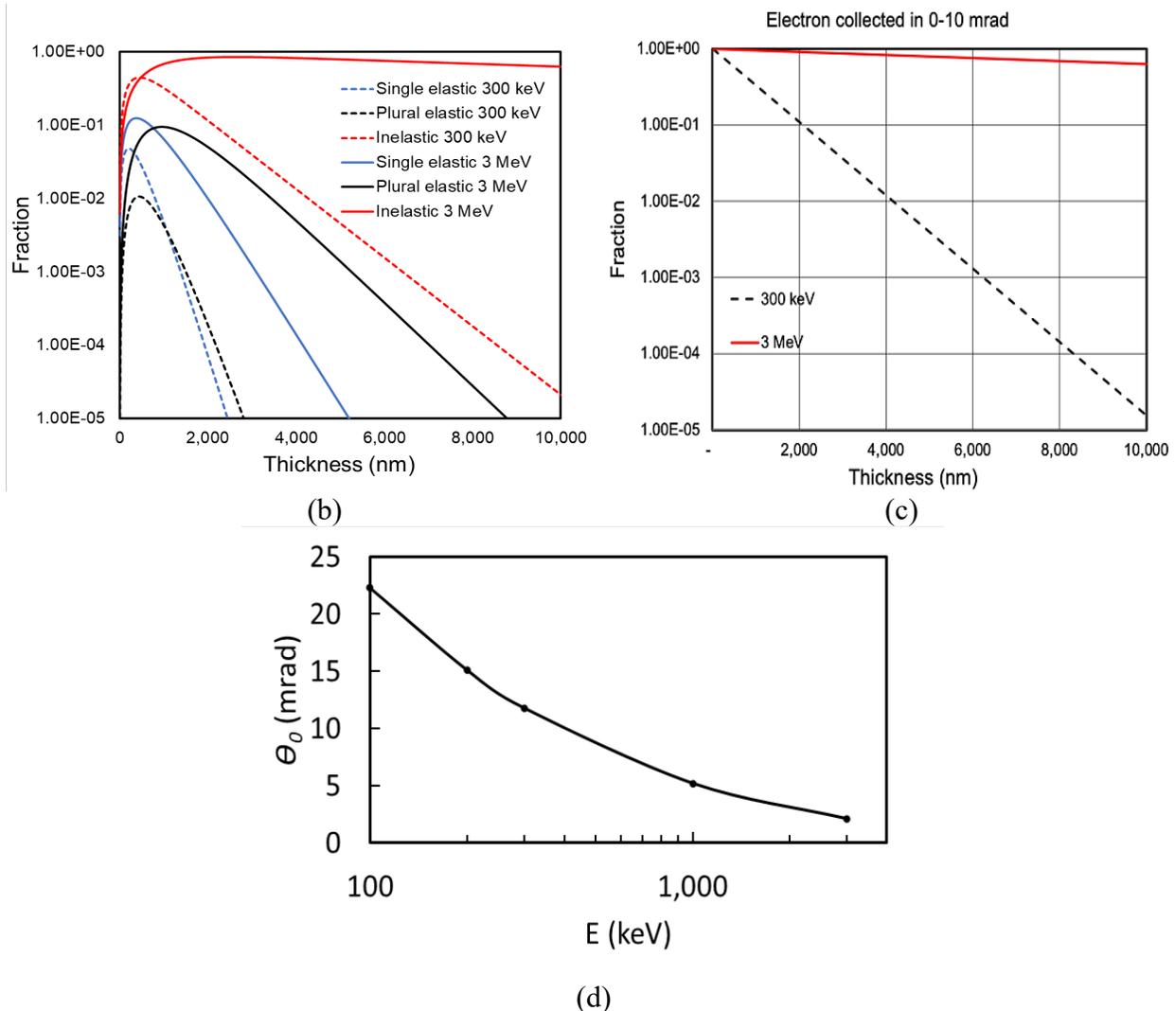

Figure 1. (a) Illustration of five electron categories ($I_{noscat}$, $I_{1el}$, $I_{el,plural}$, $I_{inel}$, $I_{out}$) after incident electron ($I_0$) and sample interaction, where $\Delta E_i$ ($\approx 40$ eV) is the single inelastic energy loss for amorphous ice and EELS means electron energy loss spectroscopy. Normalized intensity profiles for bright field STEM imaging in amorphous ice (collection angle 0-10 mrad): (b) Electrons scattered in ice as a function of thickness at incident electron energies of 300 keV (dashed lines) and 3 MeV (solid lines) including single elastic (blue), plural elastic (black), and inelastic scattering (red). (c) Total electrons collected within 0-10 mrad (the sum of all electrons shown in panel (b) decreases as the sample thickness increases). Black dash line: 300 keV electrons; red solid line: 3 MeV electrons. (d) Characteristic angle of elastic scattering ($\theta_0$) as a function of the beam energy.

### 1.4 Image contrast and resolution

In STEM, the resolution is mainly determined by the probe size and electron wavelength. The STEM probe size should be as small as possible (e.g., no more than 2 nm in diameter for the targeted resolution). As the electron probe rasters over the specimen point-by-point, the accuracy of the steering coil to shift the electron probe (scanning accuracy) should be no more than half of the probe size, 1 nm. When the sample is thick (more than 1,000 nm), the angular distribution of scattered electrons produces a spatial distribution that broadens the incident electron probe normal to the beam direction (Fig. 2a). The broadening of the probe in thick biological samples is, to first approximation, governed by the probe semi-convergence angle and the scattering of electrons in the sample. When a point electron probe is focused on the center of the specimen, the minimal geometrical broadening of the exiting electron probe $d_{bottom}$ is

$$d_{bottom} = t \cdot \tan(\alpha) \cong t \cdot \alpha, \tag{13}$$

where $t$ is the sample thickness and $\alpha$ is beam semi-convergence angle. As shown in Fig. 2, the larger the semi-convergence angle, the larger the exiting beam size.

The broadening of the probe due to electron scattering can be estimated based on a wave optical multislice algorithm [32,33]. The plural elastic scattering of electrons at 200 keV in vacuum and amorphous ice has been estimated by Wolf et al [22]. When multiple scattering is included, the broadening of the probe is expected to be worse. Based on their estimation, a schematic beam broadening for 3 MeV is sketched in Fig 2a. Assuming the probe size of 2 nm in focus, the smallest beam size exiting the sample is estimated to be from 3 to 12 nm for 1 to 10 m-thick sample. The corresponding resolution will be mainly limited by the electron beam broadening, ranging from 6 to 24 nm (The absolute resolution of a sensor is determined by its Nyquist limit, which is twice of the pixel size).

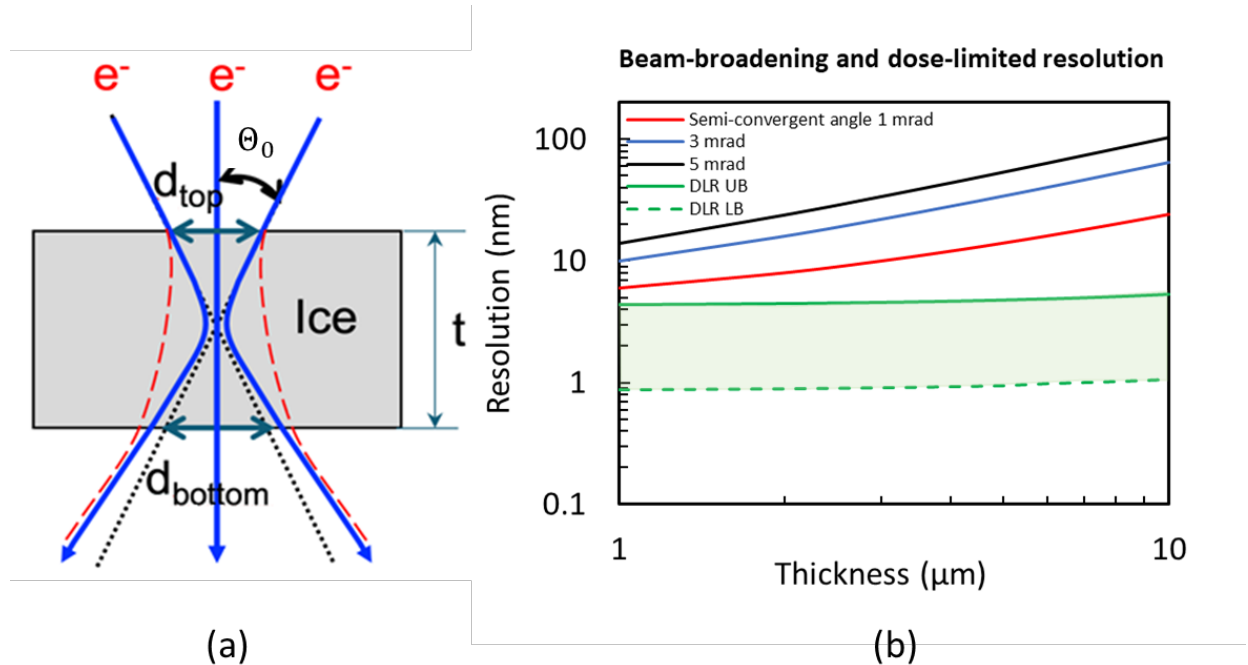

Figure 2. (a) Schematic of electron beam broadening. This represents the vacuum case with the assumption of the electron probe being focused on the center of the specimen (blue curves). The red dashed curves show the estimated broadening of the beam in amorphous ice [22]. (b) The resolution (in diameter), considering the electron beam broadening in the sample as the limiting factor, is shown as a function of the sample thickness for three different semi-convergent angles, 1 mrad (red), 3 mrad (blue) and 5 mrad (black), respectively. Also, the dose-limited resolution (DLR) is plotted as a function of the sample thickness for DLR UB (see details later) with the image contrast C=0.1 (green solid) and DLR LB with C=0.4 (green dash), respectively. Here, the image contrast C is defined as the density difference between the features and the background over the density of the background.

A dose-limited resolution (DLR) imposed by radiation damage of imaging electrons can be estimated using a standard formula [34,35]:

$$\delta_{Dose} = \sqrt{2} \cdot SNR \cdot \frac{1}{C \cdot \sqrt{DQE \cdot F \cdot Dc}}, \tag{14}$$

where SNR, C, DQE, F and Dc are the signal-to-noise ratio, contrast, detective quantum efficiency of the detector, fraction of electrons reached the detector (see Fig. 1c), and characteristic electron fluence, respectively. To be able to resolve a feature in an image, SNR needs to be 3 or higher according to Rose criterion, and we take SNR = 3 in our estimation. The average electron dose is about 2 e$^-$/Å$^2$ in each image of a cryo-ET tilting series, typically consisting of 40-60 images collected at different tilt angles as measured for frozen-hydrated samples [22,23]. For a modern detector, such as a direct electron detector, DQE = 0.5 is routinely achievable [34]. The fraction of electrons reached the detector (F) can be estimated from Fig. 1c.

In thick biological specimens, the density of different organelles and cells can exhibit significant variations. For example, the average protein density is 1.3-1.4 g/ml [36,37], the density of mitochondria is 1.19 g/ml [38], the average density of bacteria E Coli is 1.1 g/ml [39,40]. As one of the goals of MeV-STEM is to image an intact frozen cell without cutting it into thin slices, the contrast of the frozen cell needs to be calculated for the estimation of the DLR. In a frozen cell, the most noticeable and thickest organelle is the nucleus. The nucleus is a dense fibrillar network of DNA, RNA, and proteins, normally around 5-10 μm in diameter in many multicellular organisms. Its density is about 1.4 g/ml [41]. Taken the observation of the minimal presence of ice layers above the top surface and below the bottom surface of a frozen cell in standard cryo-ET studies, the contrast of biological sample *vs* amorphous ice can be estimated to be 0.4 (determined the DLR lower bound, named DLR LB) and 0.1 (determined the DLR upper bound, named DLR UB) for nucleus and cell. The DLR for a 10- m thick biological sample is 1.1 nm for nucleus. For other parts of a cell with a contrast of 0.1, the DLR for a 10- m thick biological sample is 5.3 nm.

The DLR is estimated as a function of the sample thickness for the upper bound with contrast C = 0.1 (green solid in Fig. 2b) and the lower bound with C = 0.4 (green dash), whereas the fraction of electrons received by the detector is obtained from Fig. 1c. Over the targeted sample thickness (< 10  m), the DLR is smaller than the resolution limited by the beam-broadening (red curve in Fig. 2b), and it likely won't limit the ultimate resolution of an MeV-STEM. Thus, the resolution in MeV-STEM is mainly limited by the beam semi-convergence angle and the beam size. To achieve higher resolution, the beam semi-convergence angle and beam size need to be as small as possible (e.g., 1 mrad, 2 nm in diameter).

**1.5 Electron beam parameters for 6-24 nm resolution with 1-10  m thick bio-sample**
Biological samples are sensitive to radiation damage. The commonly used electron dose for cryo-ET is 100-120 e$^-$/Å$^2$ distributed among 40-60 images collected at different tilt angles. The average electron dose is about 2 e$^-$/Å$^2$ in each image. The allowable electrons per beam position varies from 800 electrons to 5,000 electrons when the beam size changes from 2 nm to 5 nm (Table 3). To form a 1024×1024-pixel STEM image at each tilt angle, it will take 1 second with 1 μs dwelling time for each scanning point, which is comparable to current 300 kV STEMs. Thus, the fourth design parameter is (4) beam flux: 800-5,000 e$^-$/μs, which converts to the current of 0.13 to 0.8 nA.

| Beam size (nm)    | 2.0 | 3.0  | 4.0  | 5.0  |
|-------------------|-----|------|------|------|
| Electrons per spot| 800 | 1800 | 3200 | 5000 |

Table 3. Allowable electron dose per spot across varying beam size to mitigate potential harm to the biological sample.

The energy spread is determined by the criteria that the amount of the probe beam size blurring ($\Delta r$) on the sample caused by the energy spread ($\frac{\Delta E}{E}$) of incident electrons and chromatic aberration ($C_c$) should be much smaller than the probe size on the specimen required by MeV STEM (see Eq. 15).

$$\Delta r = C_c \cdot \alpha \cdot \frac{\Delta E}{E} \qquad (15)$$

Here, $\alpha$ is the aperture angle. We must design an optimal STEM column (see Beamline Optics section in detail) and constrain the electron beam energy spread [28]; thus, the ultimate probe size on the specimen is mainly determined by the electron beam emittance (the product of transverse size and semi-convergence angle at the beam waist).

In summary, to achieve nanometer spatial resolution for up to 10 μm-thick biological samples using MeV-STEM, there are important parameters to consider (see Table 4): (1) electron beam size ($d_{sam}$): 2 nm; (2) semi-convergence angle ($\alpha$): 1 mrad; (3) flux ($I$): 0.1 – 1 nA; (4) energy spread ($\Delta E/E$): $<10^{-4}$ (see later section in detail); and (5) scanning accuracy ($\Delta x_{cen}$): 1 nm.

| E   | $d_{sam}$ | $\alpha$ | I    | $\Delta E/E$ | $\Delta x_{cen}$ | $\epsilon_{geo}$ |
|-----|-----------|----------|------|--------------|------------------|------------------|
| MeV | nm        | mrad     | nA   |              | nm               | pm·rad           |
| 3.0 | 2         | 1        | 0.1-1| $<10^{-4}$   | 1                | 2                |

Table 4. Electron beam parameters for nanometer resolution MeV-STEM.

## II. CONCEPTUAL DESIGN OF MEV-STEM

Here, we will illustrate possible design and implementation of MeV-STEM, based on a photocathode gun followed by accelerating section, an aperture, a STEM column, which includes three condensed lenses, an objective (image forming lens) and scanning coils, a sample stage, and a detector. However, to meet the nm beam size and mrad semi-convergence angle on the specimen, the MeV-STEM requires an emittance of a few picometers (1,000 times smaller than currently achievable nanometer emittance), along with all other parameters listed in Table 4 being met simultaneously. We will show that the combination of an ultralow emittance electron source with the additional emittance reduction achieved using an optimized aperture can achieve these challenging parameters. This research will pave the way to construct a novel MeV-STEM, which has the potential of transforming our bioimaging capability in support of fundamental research in biological systems.

For the design of an MeV-STEM achieving a few nanometer spatial resolution, we foresee the following major challenges:

1. Obtaining a 2 nm sized electron beam with 1 mrad semi-convergence angle at the specimen in the pulse mode (electron beam stability, shape, and intensity distribution will be important).
2. Demonstrating the generation of an electron beam with ultralow emittance and energy spread (see Table 4), and low spatial-pointing jitter (a few angstrom on the sample) using a photocathode based superconducting radio frequency (SRF) gun or DC gun.
3. Delivering the electron beam current at the specimen: 0.1-1.0 nA.
4. Scanning the probe beam with the accuracy of half beam size at the specimen: 1 nm.

We focused on the design of three critical components required for the construction of the nanometer resolution MeV-STEM instrument:

1. MeV class electron source: a photocathode gun and optimized aperture can potentially deliver the electron beam with the required parameters in terms of emittance, energy spread and average current.
2. STEM column: not only demagnify the electron beam from a micron size at the aperture to 2 nm on the specimen plane but also provide the flexibility to adjust magnification over a wide range.
3. Electron beam scanning system capable of achieving a nanometer precision at the specimen.

We have performed electron beam transport simulations exploring two different approaches of the electron source for the MeV-STEM instrument: 1) DC gun, aperture, SRF cavities, and STEM column; 2) SRF gun, aperture, SRF cavities, and STEM column. Beam dynamic simulations show very promising results as they show that the condition exists for which the MeV-STEM requirements (see Table 4) can be met with both options. Also, we illustrate the design of a novel STEM column capable of demagnifying the beam size from 1 μm at the aperture to 2 nm at the specimen and varying such beam size from 2 nm to 16 μm, and a steering coil to scan the electron beam across the sample with 2 nm step size and 1 MHz repetition rate.

## III. CRITICAL COMPONENTS
### 3.1 Photocathode electron source

The forementioned specifications imply a required transverse normalized emittance $\epsilon_n \sim 10$ pm, which is presently beyond the state-of-the art that has been demonstrated in a photoinjector system, but is not outside the bounds of feasibility as we will show below. The emittance ($\epsilon_{geo}$) we mentioned earlier is the geometric emittance, which is $\gamma_0 (= \frac{E}{m_e c^2} \sim 5-8$, the Lorentz factor) times smaller than the normalized emittance $\epsilon_n$. A single electron per bunch at GHz repetition rate is sufficient to provide $> 0.1$ nA average current. This implies that the space charge affect will not be a factor in the dynamics, provided the source directly generates 10 pm emittance. To minimize the source emittance, we naturally seek to minimize both source size and photocathode intrinsic emittance/mean transverse energy (MTE). High quantum efficiency (QE) and low emittance electron beams provided by multi-alkali photocathodes make them of great interest for next-generation high-brightness photoinjectors. When operated with photon energy close to their workfunction, these photocathodes can provide electron beams suitable for ultrafast electron diffraction (UED) and/or ultrafast electron microscopy (UEM) by having a lower emittance and higher QEs compared to those of metals [42,43].

Even with a state-of-the art photoemission MTE of 35 meV as demonstrated in alkali antimonide photocathodes [42,43], 10 pm normalized source emittance requires an RMS emission size of 20 nm. However, due to the proximity of the optical diffraction limit, generating laser spots of even a few microns is a non-trivial task: photocathodes in most photoinjectors are typically operated in reflection mode in high electric fields of high voltage DC or RF guns, and the final lens of the optical imaging system for the laser to the photocathode surface cannot be located closer than tens of centimeters from the cathode surface itself, which typically yields photoemission source sizes > 10 μm in RMS. However, a multi-alkali photocathode grown onto a glass substrate allows the photocathode to be operated in the transmission mode [44,45]. A photocathode plug based on the INFN/DESY geometry equipped with a short focal lens sitting right behind the photocathode (left panel in Fig. 3) was successfully tested in an inverted geometry DC gun [45]. Initial laser spot sizes smaller than 2 μm (right panel in Fig. 3) have been obtained by focusing the laser light via a lens with very short focal length placed in the ultrahigh vacuum (UHV) environment of the gun, which is a few millimeters away from the back surface of the photocathode [44]. While nanopatterning of the photocathode either via work function spatial modulation or plasmonic enhancement [46] are viable routes to achieving emission sizes well below the optical diffraction limit of the drive laser, their compatibility with the low MTE photoemission conditions is not clear at present. As such, we opt for a simpler option: one would photoemit a dense bunch of electrons at the photocathode with the achievable minimum spot size (here we assume this to be $\sigma_x = 1$ μm), and use an aperture just downstream of the gun to clip down to 10 pm emittance and 1 electron/pulse. Space charge affect is significant in the dynamics up to the pinhole, and we ignore it downstream of the pinhole. Thanks to the large correlated divergence from space charge repulsion and intrinsic gun defocusing (called lateral position and energy chirp), in all simulations below the required pinhole diameters are in the order of tens of microns, for which sufficiently thick apertures can be manufactured via laser machining.

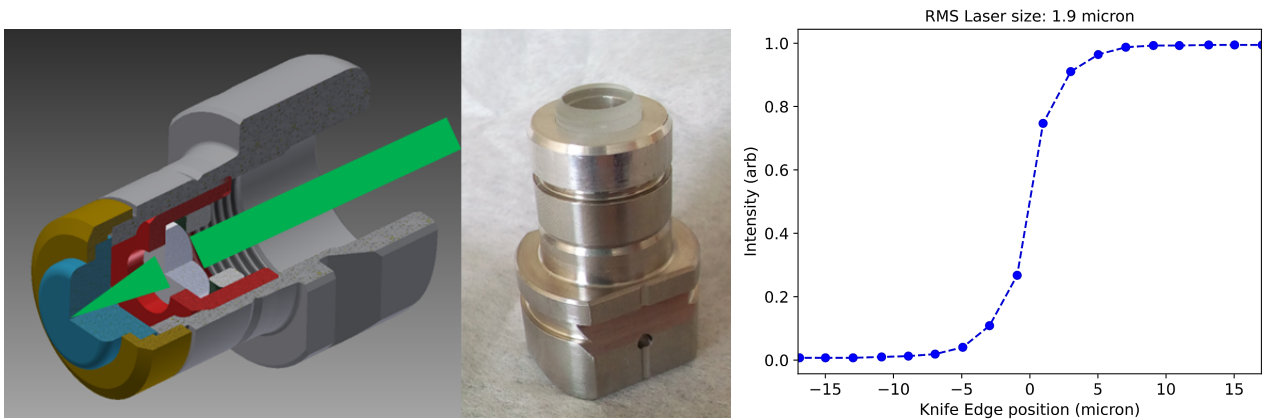

Figure 3. A modified INFN/DESY cathode plug allows microscale spot size in transmission mode with a short focal lens. Left panel: cross section view of plug concept. Center panel: Machined prototype. Right panel: knife edge beam size measurement yielding RMS spot size at photocathode surface of < 2 μm.

We choose to explore two options for the photoelectron gun: a DC (constant voltage) source and a SRF gun. The DC photogun option is to be viewed as lower risk and cost, but potentially lower performance. Several DC gun designs may be applicable here, including both traditional insulator [47] and compact inverted insulator [48] designs. DC guns delivering kinetic energy < 400 keV with photocathode electric field of ~5 MV/m are capable of reliable operation for many years. The DC gun we use in this study is the Cornell ERL photoinjector DC source, operated at 300 keV.

SRF guns promise much higher output kinetic energy and higher photocathode source field, and are compatible with low emittance semiconductor photocathodes, and thus higher brightness performance is expected. Aside from the DC gun, efforts are in place to design and demonstrate the operation of photocathode in transmission mode in SRF gun [49]. However, interfacing replaceable photocathodes with some geometries of SRF guns is a complex engineering challenge; optimizing this interface is an active area of research [49,50]. The SRF gun we choose for this study is the 1.3 GHz, 1.5 cell SRF photogun under development by Euclid [50], with output kinetic energy of (> 1.6 MeV) at 20 MV/m at the cathode, ultimately reaching 47 MV/m during experiment in liguid helium. Progress on the SRF technology based on $Nb_3Sn$ resulted in the development of accelerating cavities that may in near future be operated without the need of a cryoplant for the production of liquid helium [51]. $Nb_3Sn$ has a critical temperature of 18 K, which makes them compatible with commercially available cryostats. The case of the SRF gun built by Euclid [52] is shown in Fig. 4a. The limited cooling power provided by cryostats may complicate the use of removable photocathode plugs coated with alkali antimonide, as the stalk will likely provide yet another channel of thermal losses. But for the main purpose of our study, we will assume from here forward that similar performances from the photocathode (minimum laser spot size and photoemission MTE) can be achieved in both injector configurations.

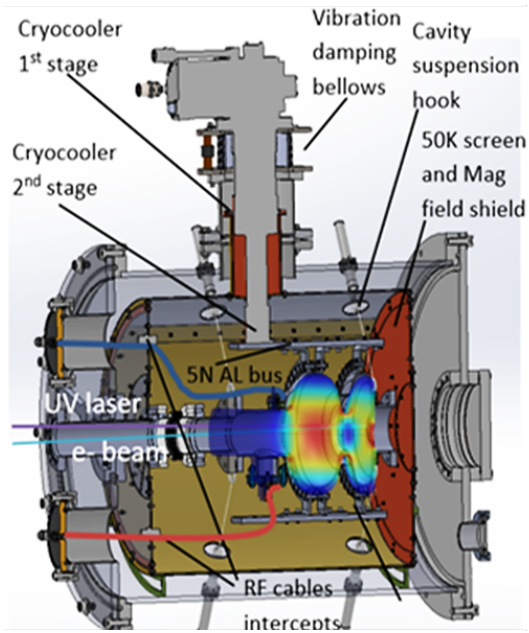
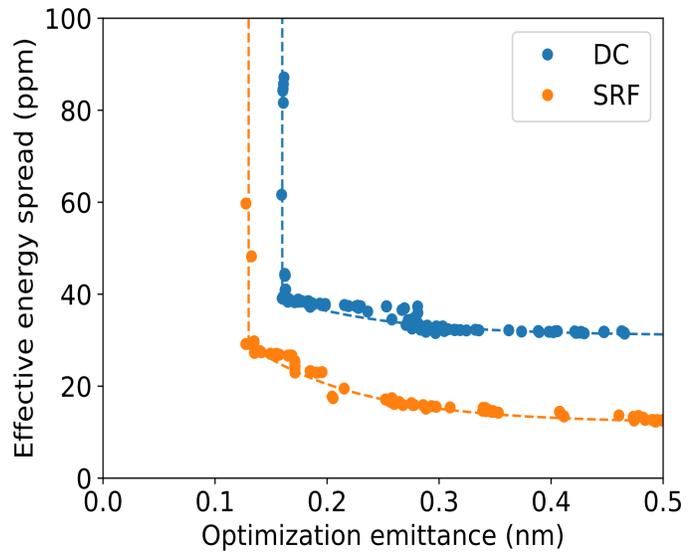

(a)                (b)

Figure 4. (a) L-band SRF photocathode gun with cryocooler built by Euclid [50]. (b) Pareto optimal performance of effective energy spread and transverse emittance for a transmitted charge of 100 aC, for both DC (blue) and SRF (orange) guns. Dotted lines guide the eye. Individual solutions from these sets are chosen and rerun with clipping down to the level of one electron per pulse on average.

Thanks to the tightly focused laser spot size of a few microns when the photocathode operated in transmission mode, the spatial pointing stability of the electron source should be better than a few percent of the laser spot size (< 100 nm) [53]; due to the factor of 500 or larger demagnification, the stability of the probe beam on the specimen will be in the level of a few angstroms.

### 3.2 Photoinjector layout and optimization

For the general photoinjector design, we take inspiration from the Cornell ERL photoinjector, which was recently considered as a source for ultrafast electron diffraction [54]. The photoelectron gun (either DC or SRF) is followed by a transversely focusing solenoid, a normal conducting 1.3 GHz bunching cavity, a second transversely focusing solenoid, and two 1.3 GHz SRF accelerating cavities. This layout is shown in Fig. 5a. We utilize multiobjective genetic optimization of space charge simulations carried out in General Particle Tracer [55,56]. We seek to simultaneously optimize emittance and energy spread from the injector downstream of the SRF cavities. Our simulations begin with an initial bunch charge of 12,500 electrons/pulse, and ultimately an aperture located within the first solenoid lens will reduce the bunch charge down to 1 electron/pulse on average. To alleviate long computation times, we first optimize with a larger pinhole which is tuned on a case-by-case basis to transmit 600 electrons/pulse ($600 \cdot e \approx 100$ aC). We will later select from the Pareto optimal front individual cases to rerun with a larger number of macroparticles to enable heavier aperture clipping. In cases where we clip down to on average 1 electron per pulse, we ignore Coulomb interaction effects after the pinhole. In reality, Poisson statistics will dicate that the number of electrons in a bunch will vary, and this fluctuating small number of electrons will interact via the bare Coulomb potential; this stochasticity is beyond the scope of this work and is left for future study.

The optimizer is permitted to modify the laser duration (constrained to be $\sigma_t > 1$ ps), and the RMS tansverse laser size ($\sigma_x > 1$ µm), where the longitudinal distribution is assumed to be uniform and the transverse distribution is assumed to be fully Gaussian. The photoemission MTE is set to be 35 meV. The optimizer has the freedom to change the amplitudes and phases of all fields within practical bounds. As was done for temporal resolution in a previous work [54], our figure of merit for energy spread includes both the energy spread of the bunch and effects of energy jitter. Using previously characterized values for amplitude and phase stability of the accelerating elements, we compute multiple simulations per machine setpoint within the bounds of this jitter to determine the RMS energy jitter value $\sigma_{E,j}$. The effective energy spread ($\sigma_{E,eff}$) we compute is the quadrature sum of the bunch energy spread $\sigma_E$ and the RMS energy jitter $\sigma_{E,j}$. The total kinetic energy of the bunch after acceleration is constrained to lie within 2.5 and 3.5 MeV.

The Pareto optimal frontier of energy spread and emittance with a final charge of 100 aC/pulse is shown in Fig. 4b. The SRF gun is capable of outperforming the DC gun in both emittance and effective energy spread. The emittance improvement we attribute to the use of higher photocathode field, which permits higher charge densities on the cathode, and relativistic energies which suppresses the space charge interaction force. The energy spread improvement is primarily due to less energy jitter, as amplitude and phase changes in the gun lead to smaller arrival time differences at the downstream cavities since the bunch velocity at the gun exit is nearly saturated at $c$ (speed of light). In Figs. 5b to 5g, we have selected optimal solutions for both DC and SRF guns, and have rerun them for the case of clipping down to one electron per pulse on average. Emittance values in both cases are very near 10 pm with total effective energy spread on the scale of 100 eV, which suggests both beamline designs are appropriate sources for MeV-STEM.

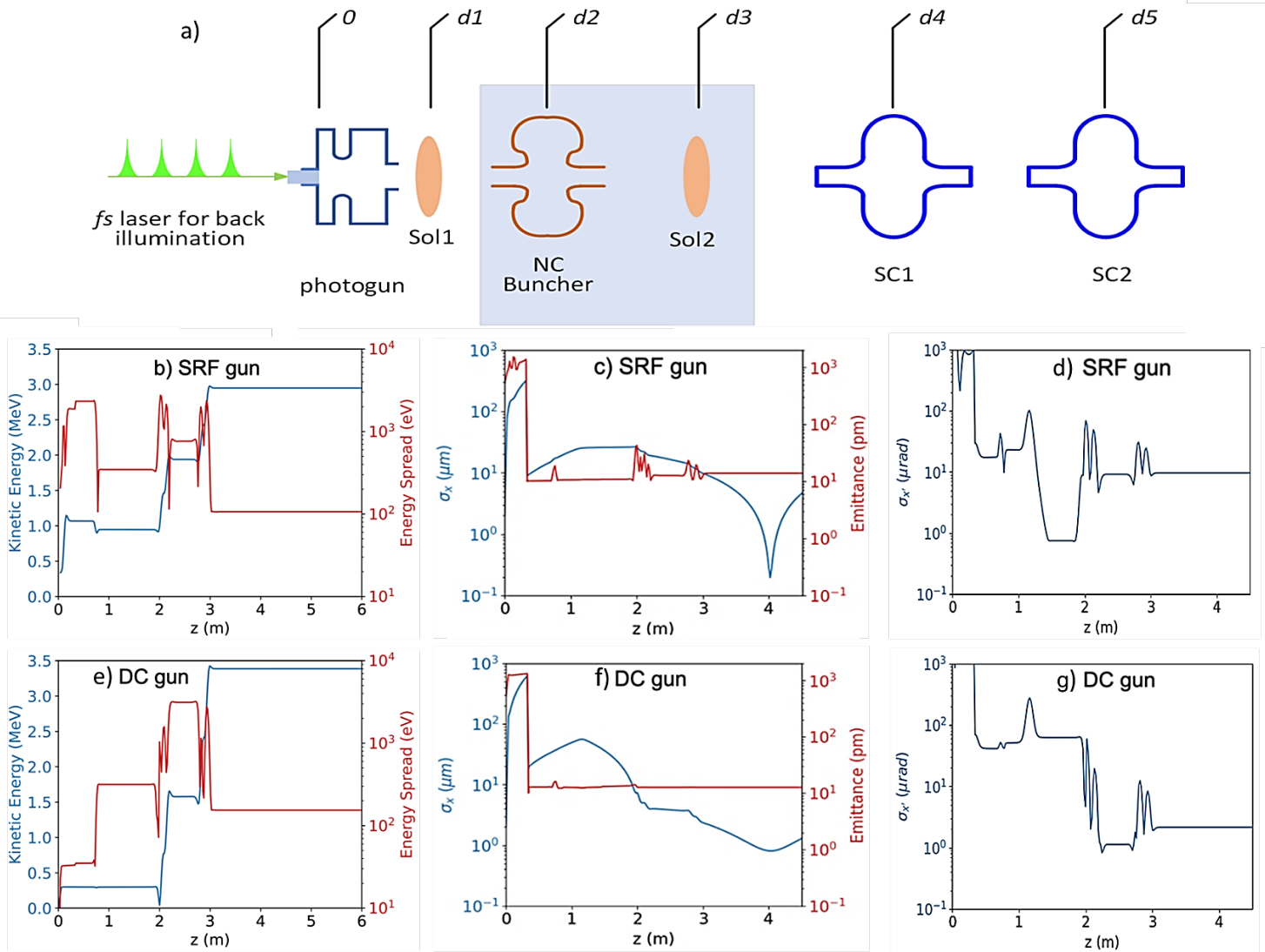

Figure 5. (a) Beamline layout showing the longitudinal positions of the photoelectron gun (either DC or SRF), and the common section after the gun : transverse focusing solenoids, a normal conducting 1.3 GHz buncher cavity, and two 1.3 GHz SRF cavities. An aperture within the first solenoid is used to reduce emittance down to about 10 pm. The blue colored area highlights that in the SRF gun case, the section from the exit of first solenoid to the exit of second solenoid can be eliminated without any degradation of the electron source properties; in the DC gun case, this section is still needed to keep similarly minimized emittances and energy spread. Here, d1=0.35 m, d2=0.75 m, d3=1.15 m, d4=2.10 m, and d5=2.85 m are the longituidnal positions of solenoid 1, buncher cavity, solenoid 2, SRF accelerating cavity 1 and 2, respectively. (b), (c) and (d) Energy, energy spread, transverse size, transverse emittance, and transverse divergence for the SRF gun case. (e), (f) and (g) Same but for the DC gun case. Transverse emittance remains constant after z = 1 m.

### 3.3 Beamline Optics: STEM Column

We adopted a similar design based on permanent quadrupole (PMQ) quintuplets pioneered at Brookhaven National Laboratory for MeV electron microscopes [28] to construct the MeV-STEM column, since the electron optics of a STEM column is similar to that of a TEM column based on reciprocity [57]. The STEM column is used not only to focus the electron beam onto the sample with the size of a few nanometers, but also to provide the beam with various sizes at the specimen to enable scanning at different magnifications. We used four lenses in the STEM column, with one objective lens and three condenser lenses (see Appendices and Fig. 6a for details). To keep the minimum aberrations, the objective lens has to be much stronger than the condenser lenses. As a result, the pole tip radius of the objective lens is half of the value of the condenser lenses and ultimately limits the maximum probe size that can be achieved on the sample. The design strategies are: 1) the strength of each lens can be adjusted to vary the probe size 1400 and 5360 times for the SRF and DC gun cases; 2) the aperture in solenoid 1 provides the additional freedom of changing the electron beam size at the exit of the electron source up to 10 and 2.5 times for the SRF and DC gun case, respectively (see Figs. 5c and 5f for the aperture at $z = d1$), hence the ultimate probe size at the specimen can be varied from 2 nm to 16 µm. The variation of the aperture can greatly mitigate the required change of the magnetic field of those lenses to ±8.2 T/m and ±10.8 T/m for the SRF and the DC gun cases, respectively. For 1 mm pole tip radius, ±8.2 T/m and ±10.8 T/m correspond to ±3.3 Ampere turn and ±4.3 Ampere turn, respectively. This entails that it is feasible to add coils to the PMQs to adjust their field strengths to achieve the required change in probe size.

From Figs. 5c and 5f, we learn that the RMS size of the electron beam and normalized emittance at the end of the source (z = 4 m) for the cases of the SRF gun and the DC gun are 0.2 µm and 0.8 µm, respectively. The normalized emittances for the cases of the SRF gun and the DC gun are 13.9 pm and 13.1 pm, respectively. The kinetic energies of the electrons at the end of SRF and the DC source are 3.0 MeV and 3.4 MeV, respectively. Lastly, the opening angles at the sample are 10.0 µrad and 2.2 µrad, respectively. Because the beam sizes at the exit of the two sources differ by a factor of four, the necessary demagnification of the columns also differs by a factor of four. The precise values of the demagnifications will be determined below, minimizing the beam size at the sample based on the aberrations of the objective lens. In order to minimize the aberrations, the focal length of the objective lens is minimized through reducing the pole tip radius (allowing for higher gradient), the lengths of the magnets and the spacing between them. The results are listed in Table S1 in Appendices (see Fig. 7a for magnetic field profile). Since the aberrations of the condenser lens don't affect the resolution significantly, the focal length can be larger, which greatly relaxes the design requirement of the condenser lens (in Appendices, see Table S2 for condenser lenses 1 and 2 and Table S3 for condenser lens 3). The layout of the STEM column is shown in Table 5. The aberrations of the objective lens parameters are shown in Table 6.

| Element | L (SRF/DC, mm) | σ (SRF, μm) | σ (DC, μm) |
|---|---|---|---|
| Electron Source Exit | 0 | 0.2 – 2 | 0.8 – 2 |
| Drift ($L_1$) | 20 | 0.29 – 2.9 | 0.8 – 2 |
| Condenser Lens ($C_1$) | 76 | 0.48 – 5.8 | 0.1 – 0.6 |
| Drift ($L_2$) | 500 | 1.76 – 17.6 | 6.9 – 17.3 |
| Condenser Lens ($C_2$) | 76 | 0.36 – 6.1 | 0.02 – 5.9 |
| Drift ($L_3$) | 500 | 14.6 – 146 | 57.2 – 145 |
| Condenser Lens ($C_3$) | 76 | 5.0 – 50 | 19.6 – 124 |
| Drift ($L_4$) | 350 | 18.1 – 830 | 70.9 – 814 |
| Objective Lens (O) | 17 | 0.7 – 15.3 | 2.6 – 15.0 |
| Drift ($L_5$) | 0.17 | 0.002 – 16 | 0.002 – 16 |

Table 5. Layout of the STEM column ($L_{tot} = 1.615\ m$).

|  | f (cm) | $C_{s,x}$ (cm) | $C_{s,y}$ (cm) | $C_{s,xy}$ (cm) | $C_{c,x}$ (cm) | $C_{c,y}$ (cm) |
|---|---|---|---|---|---|---|
| Probe forming Lens (SRF) | 1.00 | 16.2 | 15.0 | 35.3 | 1.78 | 1.44 |
| probe forming Lens (DC) | 1.00 | 16.2 | 15.0 | 35.3 | 1.80 | 1.46 |

Table 6. Objective lens parameters including the focal length and main aberrations (obtained using the hyper-tangent fringe-field model).

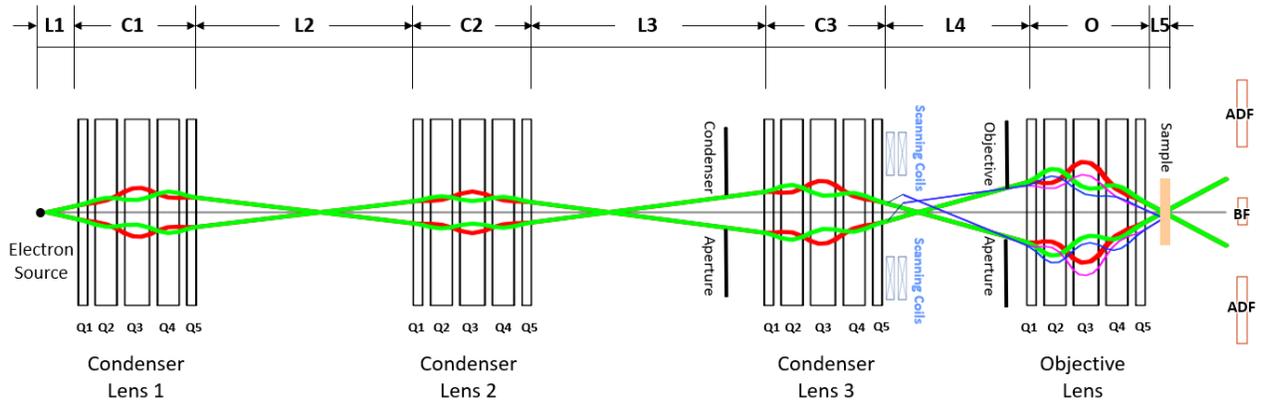

(a)

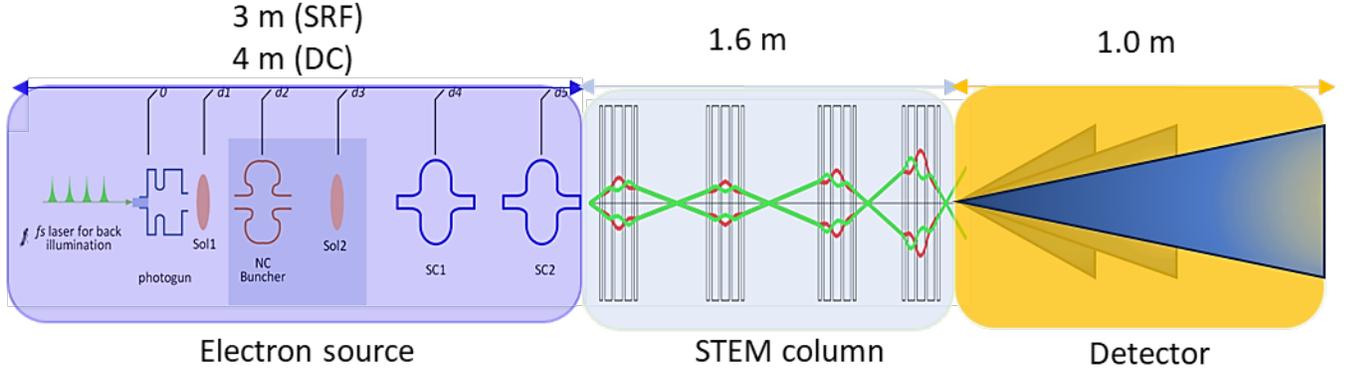

(b)

Figure 6. (a) Layout of the STEM column based on the quadrupole quintuplet lenses. The red and green ray diagrams are the calculated beam path in the x-z and y-z direction, respectively. (b) Layout with dimensions of the MeV-STEM instrument.

In order to obtain the final probe size, the effect of the aberrations of the objective lens has to be included, which are shown in Table 6. From Figs. 5b and 5e, energy spread for SRF and DC sources are roughly 107 eV and 155 eV, respectively. The relative energy spread are $\Delta E/E = 3.3 \times 10^{-5}$ and $4.4 \times 10^{-5}$, respectively. The final probe beam size in RMS is

$$\sigma_{tot} = \sqrt{\sigma_d^2 + \sigma_s^2 + \sigma_c^2 + \sigma_{emit}^2} \tag{16}$$

where $\sigma_d = 0.61\lambda/\alpha$, $\sigma_s = C_s\alpha^3/4$, taking into account the disk of least confusion, and $\sigma_c = C_c\alpha\Delta E/E$, and $\sigma_{emit} = \epsilon_{geo}/\alpha$. Parameters $\lambda$, $\alpha$, $C_S$, $C_C$, $\Delta E/E$, and $\epsilon_{geo}$ are the wavelength, the aperture angle on the sample, the spherical aberration, the chromatic aberration, the relative energy spread and the geometrical transverse emittance, respectively. The wavelengths for the SRF and the DC sources are 0.36 pm and 0.32 pm, respectively. The theoretically minimum resolution can be obtained by

$$\delta_{th} = \sqrt{\sigma_d^2 + \sigma_s^2 + \sigma_c^2}. \tag{17}$$

The result is shown in Fig. 7b. Unlike conventional STEM, the resolution of this instrument is limited by the chromatic aberration due to its larger energy spread. The theoretically minimum resolutions for the SRF (aperture angle 0.58 mrad) and the DC (aperture angle 0.49 mrad) sources are 0.53 nm and 0.57 nm, respectively. If the current is of no concern, the best theoretical resolution of the instrument can always be reached through increasing demagnification, resulting in a small probe size. Yet in this design, at roughly 13 to 14 pm normalized emittance, there is on average only 1 electron per pulse, which entails that the maximum current is around 200 pA. Thus, the emittance of the electron beam can't be much further reduced. The effect of the emittance on the resolution is shown in Fig. 7b. The minimum resolution values for the SRF and the DC cases are 1.63 nm and 1.69 nm, respectively, with aperture angle 1.75 mrad and 1.44 mrad, respectively. The resulted demagnifications for the SRF and the DC cases are 175 and 670, respectively. The demagnification of the objective lens is chosen to be 20 for both cases. The demagnifications of the condenser lens for the SRF and the DC cases are 8.75 and 33.5, respectively.

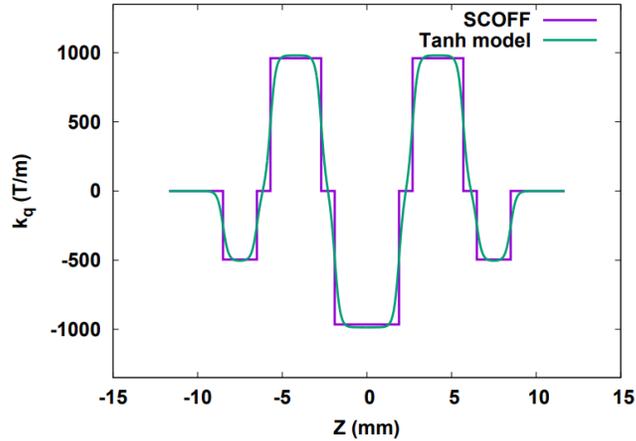

(a)

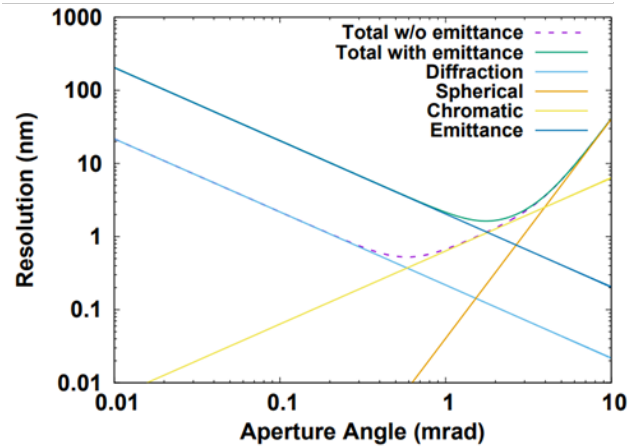

(b)

Figure 7. (a) Magnetic field profile of the objective lens. (b) RMS resolution for the SRF gun (That for the DC gun is not shown since it looks rather similar.). The "Total w/o emittance" is the resolution defined in Eq. (17). The "Total with emittance" is the resolution defined in Eq. (16).

### 3.4 Scanning coils

To demonstrate the feasibility of rastering an nm-sized probe beam across the specimen with a sub-nm precision, we numerically study the critical component – scanning, or steering coils. Based on the preliminary design of the MeV-STEM instrument, simulation is carried out in GPT. To cover the sample with 2000×2000 pixels at the step size of 2 nm, scanning coils need to be positioned right after the last condenser lens. A numerical study indicates that the entire range of the steering should cover ±0.32 mrad with a step size of 160 nrad. The 0.2 nm scanning accuracy (10% beam size) requires a power supply with a precision better than 25 ppm, which is achievable based on the currently demonstrated performance of power supplies.

A schematic layout of the MeV-STEM instument is shown in Fig. 6b. Despite the two options based on either a DC gun or a SRF gun, the common section of the MeV-STEM instrument starts from the aperture and ends with the detector. Also, the same drive laser system can be applied to both the DC and SRF guns. The photocathode drive laser can be a turn-key femtosecond laser with a 1 to 10 GHz repetition rate, delivering more than 1 W of average power in pulses as short as 50 fs, from Novanta Photonics [58]. The pulse energy of 1 nJ can generate the charge >2 fC per pulse with the multi-alkali photocathode, whereas only 2 fC/pulse is needed.

**METHOD**

A promising alternative for imaging thick biological samples (e.g., eukaryotic cells, thick section of tissue and organs) is to employ STEM, where the electron beam is focused on the specimen and rastered across it point by point. In STEM, there are no image-forming lenses after the specimen. An image is formed by mapping the detector counts with each scanning position. Thus, STEM image-formation is an incoherent process, which is less affected by inelastic scattering. Instead of forming an image via coherent interference (i.e., elastic scattering) among the electron waves in cryo-EM or cryo-ET, the intensity is directly summed together, and inelastic scattered electrons contribute to the signal. Thus, STEM imaging can be very effective for thick samples. As demonstrated [22-23], STEM has been employed to study biological samples up to 1,000 nm thick instead of 300 nm thick for cryo-EM and cryo-ET in 200-300 keV TEM [22].

To achieve 6-24 nm resolutions with 1-10 m thick biological samples, an MeV-STEM requires an electron source with the emittance of a few picometer, which is ~1,000 times smaller than the presently achieved nm emittance, in conjunction with less than $10^{-4}$ energy spread and 1 nA current. We numerically demonstrated the feasibility of building such an MeV-STEM instrument and applying it to study large/thick bio-samples with the thickness up to 10 μm.

**DISCUSSION**

As the allowable sample thickness without radiation damages depends on the electron energy and image formation mechanism, we propose to develop a MeV-STEM. The remarkable ability of MeV-STEM to penetrate deeply into samples, even as thick as 10 μm or more, while still achieving nanoscale resolution, positions it as an excellent choice for biological specimen analysis. The impact of electron energy, beam broadening, and low-dose constraints on resolution has been examined. Notably, the finest achievable resolution exhibits an inverse relationship with sample thickness, shifting from 6 nm for 1 μm thick samples to 24 nm for 10 μm thick ones. These make MeV-STEM capable of imaging large and thick biological sample (up to 10 m in thickness). However, achieving MeV-STEM capabilities necessitates electron beam emittance that goes beyond the state-of-the-art (a few pm emittance, better than $10^{-4}$ energy spread and 0.1-1 nA current).

Our preliminary simulation study as well as our MeV-UEM hardware research and development suggest building such an MeV-STEM is possible. A photocathode gun based on the ERL injector at Cornell University can produce electron beams with a few picometer emittance, better than $10^{-4}$ energy spread, and nA beam current. Since all the critical components (ultra-low emittance DC gun, SRF accelerating cavity and momentum aperture) already have been demonstrated and reliably operated, the MeV-STEM source can be realized based on this design. Alternatively, a compact SRF gun similar to the one built by Euclid Techlabs may provide even better performances for an MeV-STEM instrument. In the last case several significant challenges related to the cathode insertion mechanism and its operation in transmission mode have yet to be addressed before the last solution can be adopted. So far, it has been numerically confirmed that in the SRF gun case, one can eliminate the section including the buncher cavity and second solenoid while keeping similarly minimized emittance and energy spread. The MeV-STEM instrument can be built within a total length less than 6 m.

Furthermore, the numerically demonstrated ultralow emittance ($\epsilon_{geo}$=2 pm) MeV electron source can deliver a quasi- monochromatic pencil beam. Such MeV electron source could help in experimentally answering some long-standing challenging problems as well as easily be reconfigured into single-shot MeV-UED/UEM operational mode without any hardware change, which is highly desired for solving challenges in probing matter at ultrafast-temporal ultrasmall-spatial scales.

## ACKNOWLEDGMENTS
We would like to thank Dr. Sharon Wolf for her valuable insights and discussions on using STEM for imaging biological samples, Drs. Ming Du and Chris J Jacobsen for their enlightening discussions on electron scattering in specimens, and Dr. Ray Egerton for discussion of image resolutions.

This work was partially supported by the U.S. Department of Energy under grant DE-SC0012704 and BNL LDRD 22-029.

## COMPETING INTERESTS
The authors declare no competing financial and non-financial interests in relation to the work described in the paper.

## AUTHOR CONTRIBUTIONS
L. Wang introduced the concept of MeV-STEM for imaging thick bio-samples, conducted calculations of the intensity change of electrons in various categories while traveling through thick samples, estimated the contrast in biological samples, and analyzed the effect of beam broadening and low-dose limit on resolution. X. Yang proposed technical solution for MeV electron sources, specifically the DC and SRF photo-guns, to meet the requirement of nanometre resolution MeV-STEM, formulated the ultimate resolution including source emittance, and made critical suggestion on extending the adjustable magnification of the STEM column. W. Wan was responsible for the design of the STEM column, especially for finding the optimal solution of STEM column with the minimum aberration and maximum tuning range. J. Maxson, A. Bartnik, and M. Kaemingk performed beam dynamic simulations and optimized the performance of the electron source. C. Jing and R. Kostin provided the field map of the SRF photocathode gun. L. Cultrera contributed to the design of the photocathode. Y. Zhu provided critical views on the complexity of electron scattering in thick samples and its impact on resolution and offered guidance and expertise on the design of STEM column. L. Wu for discussions and dynamic calculations of elastic scattering for thick samples. S. McSweeney provided advice on the design of MeV-STEM, T. Shaftan and V. Smaluk provided advice on design issues of the instrument and overall structure of the paper. All authors actively participated in figure preparation and collectively contributed to the writing of the manuscript.

## DATA AVAILABILITY

The datasets generated and analyzed during the current study are not publicly available due to the reason that we want to know who has an interest in our datasets but are available from the corresponding author upon reasonable request.

## REFERENCE

1. Tegunov, D., Xue, L., Dienemann, C., Cramer, P. & Mahamid, J. Multi-particle cryo-EM refinement with M visualizes ribosome-antibiotic complex at 3.5 Å in cells. *Nat. Methods* **18**, 186-193, doi:10.1038/s41592-020-01054-7 (2021).
2. Nicastro, D. *et al.* The molecular architecture of axonemes revealed by cryoelectron tomography. *Science* **313**, 944-948, doi:10.1126/science.1128618 (2006).
3. Dobro, M. J. *et al.* Uncharacterized Bacterial Structures Revealed by Electron Cryotomography. *J. Bacteriol.* **199**, doi:10.1128/jb.00100-17 (2017).
4. Li, X. Cryo-electron tomography: observing the cell at the atomic level. *Nat. Methods* **18**, 440-441, doi:10.1038/s41592-021-01133-3 (2021).
5. Brocard, L. *et al.* Proteomic Analysis of Lipid Droplets from Arabidopsis Aging Leaves Brings New Insight into Their Biogenesis and Functions. *Front Plant Sci* **8**, 894, doi:10.3389/fpls.2017.00894 (2017).
6. Jin, X. *et al.* Three-Dimensional Analysis of Chloroplast Structures Associated with Virus Infection. *Plant Physiol.* **176**, 282-294, doi:10.1104/pp.17.00871 (2018).
7. Yi-Wei, C. *et al.* Architecture of the type IVa pilus machine. *Science (American Association for the Advancement of Science)* **351**, 1165-1165, doi:10.1126/science.aad2001 (2016).
8. Weiner, E., Pinskey, J. M., Nicastro, D. & Otegui, M. S. Electron microscopy for imaging organelles in plants and algae. *Plant Physiol* **188**, 713-725, doi:10.1093/plphys/kiab449 (2022).
9. Böhning, J. & Bharat, T. A. M. Towards high-throughput in situ structural biology using electron cryotomography. *Prog Biophys Mol Biol* **160**, 97-103, doi:10.1016/j.pbiomolbio.2020.05.010 (2021).
10. Oikonomou, C. M. & Jensen, G. J. Cellular Electron Cryotomography: Toward Structural Biology In Situ. *Annu Rev Biochem* **86**, 873-896, doi:10.1146/annurev-biochem-061516-044741 (2017).
11. Otegui, M. S. & Pennington, J. G. Electron tomography in plant cell biology. *Microscopy (Oxf)* **68**, 69-79, doi:10.1093/jmicro/dfy133 (2019).
12. Cooper, C., Thompson, R. C. A. & Clode, P. L. Investigating parasites in three dimensions: trends in volume microscopy. *Trends in Parasitology* **39**, 668-681, doi:https://doi.org/10.1016/j.pt.2023.05.004 (2023).
13. Collinson, L. M. *et al.* Volume EM: a quiet revolution takes shape. *Nature Methods* **20**, 777-782, doi:10.1038/s41592-023-01861-8 (2023).
14. Capua-Shenkar, J. *et al.* Examining atherosclerotic lesions in three dimensions at the nanometer scale with cryo-FIB-SEM. *Proc Natl Acad Sci U S A* **119**, e2205475119, doi:10.1073/pnas.2205475119 (2022).
15. Schertel, A. *et al.* Cryo FIB-SEM: Volume imaging of cellular ultrastructure in native frozen specimens. *Journal of Structural Biology* **184**, 355-360, doi:https://doi.org/10.1016/j.jsb.2013.09.024 (2013).
16. Spehner, D. *et al.* Cryo-FIB-SEM as a promising tool for localizing proteins in 3D. *Journal of Structural Biology* **211**, 107528, doi:https://doi.org/10.1016/j.jsb.2020.107528 (2020).

17  Vidavsky, N. *et al.* Cryo-FIB-SEM serial milling and block face imaging: Large volume structural analysis of biological tissues preserved close to their native state. *Journal of Structural Biology* **196**, 487-495, doi:https://doi.org/10.1016/j.jsb.2016.09.016 (2016).
18  Raguin, E., Weinkamer, R., Schmitt, C., Curcuraci, L. & Fratzl, P. Logistics of Bone Mineralization in the Chick Embryo Studied by 3D Cryo FIB-SEM Imaging. *Adv Sci (Weinh)* **10**, e2301231, doi:10.1002/advs.202301231 (2023).
19  Du, M. & Jacobsen, C. Relative merits and limiting factors for x-ray and electron microscopy of thick, hydrated organic materials. *Ultramicroscopy* **184**, 293-309, doi:10.1016/j.ultramic.2017.10.003 (2018).
20  Wolf, S. G. & Elbaum, M. CryoSTEM tomography in biology. *Methods Cell Biol* **152**, 197-215, doi:10.1016/bs.mcb.2019.04.001 (2019).
21  Hohmann-Marriott, M. F. *et al.* Nanoscale 3D cellular imaging by axial scanning transmission electron tomography. *Nature Methods* **6**, 729-731, doi:10.1038/nmeth.1367 (2009).
22  Wolf, S. G., Shimoni, E., Elbaum, M. & Houben, L. in *Cellular Imaging: Electron Tomography and Related Techniques*   (ed Eric Hanssen)  33-60 (Springer International Publishing, 2018).
23  Wolf, S. G., Houben, L. & Elbaum, M. Cryo-scanning transmission electron tomography of vitrified cells. *Nat Methods* **11**, 423-428, doi:10.1038/nmeth.2842 (2014).
24  Hawkes, P. W.  Vol. 159   (Elsevier, 2009).
25  Reimer, L. & Kohl, H. *Transmission electron microscopy : physics of image formation*. 5th edn, (New York, NY : Springer, c2008, 2008).
26  Tanaka, N. *Electron Nano-Imaging Basics of Imaging and Diffraction for TEM and STEM*. 1st edn, (Springer, Tokyo. , 2017).
27  Dupouy, G. Performance and applications of the Toulouse 3 million volt electron microscope. *J Microsc* **97**, 3-28, doi:10.1111/j.1365-2818.1973.tb03757.x (1973).
28  Wan, W., Chen, F. R. & Zhu, Y. Design of compact ultrafast microscopes for single- and multi-shot imaging with MeV electrons. *Ultramicroscopy* **194**, 143-153, doi:10.1016/j.ultramic.2018.08.005 (2018).
29  Tanaka, N. *Electron Nano-Imaging*. 1 edn,  (Springer, Tokyo, 2017).
30  Williams , D. B. & Carter, C. B. *Transmission Electron Microscopy*.  (Springer New York, NY, 2009).
31  Jacobsen, C., Medenwaldt, R. & Williams, S. in *X-Ray Microscopy and Spectromicroscopy: Status Report from the Fifth International Conference, Würzburg, August 19–23, 1996*   (eds Jürgen Thieme, Günter Schmahl, Dietbert Rudolph, & Eberhard Umbach)  197-206 (Springer Berlin Heidelberg, 1998).
32  Wu, L. *et al.* Valence-electron distribution inMgB2by accurate diffraction measurements and first-principles calculations. *Physical review. B, Condensed matter and materials physics* **69**, doi:10.1103/PhysRevB.69.064501 (2004).
33  Li, J. *et al.* Dichotomy in ultrafast atomic dynamics as direct evidence of polaron formation in manganites. *npj Quantum Materials* **1**, 16026, doi:10.1038/npjquantmats.2016.26 (2016).
34  Egerton, R. F. Choice of operating voltage for a transmission electron microscope. *Ultramicroscopy* **145**, 85-93, doi:10.1016/j.ultramic.2013.10.019 (2014).
35  Reimer, L. & Kohl, H. *Transmission Electron Microscopy*. 5th edn,  (Springer, New York, 2008).
36  Durchschlag, H. in *Thermodynamic Data for Biochemistry and Biotechnology*   (ed Hans-Jürgen Hinz)  45-128 (Springer Berlin Heidelberg, 1986).
37  Howard, J. *Mechanics of Motor Proteins and the Cytoskeleton*.  30 (Sinauer Associates, Inc, 2001).

38    Rickwood, D., Chambers, J. A. A. & Barat, M. Isolation and preliminary characterisation of DNA-protein complexes from the mitochondria of Saccharomyces cerevisiae. *Experimental Cell Research* **133**, 1-13, doi:https://doi.org/10.1016/0014-4827(81)90350-5 (1981).

39    Baldwin, W. W., Myer, R., Powell, N., Anderson, E. & Koch, A. L. Buoyant density of Escherichia coli is determined solely by the osmolarity of the culture medium. *Arch Microbiol* **164**, 155-157, doi:10.1007/s002030050248 (1995).

40    Loferer-Krössbacher, M., Klima, J. & Psenner, R. Determination of bacterial cell dry mass by transmission electron microscopy and densitometric image analysis. *Appl Environ Microbiol* **64**, 688-694, doi:10.1128/aem.64.2.688-694.1998 (1998).

41    Wang, N. S. *Cell fractionation based on density gradient*, <https://user.eng.umd.edu/~nsw/ench485/lab10.htm> (2023).

42    Li, W. H. *et al.* A kiloelectron-volt ultrafast electron micro-diffraction apparatus using low emittance semiconductor photocathodes. *Structural Dynamics* **9**, doi:10.1063/4.0000138 (2022).

43    Gordon, M. *et al.* Four-dimensional emittance measurements of ultrafast electron diffraction optics corrected up to sextupole order. *Physical Review Accelerators and Beams* **25**, 084001, doi:10.1103/PhysRevAccelBeams.25.084001 (2022).

44    Lee, H., Cultrera, L. & Bazarov, I. Intrinsic emittance reduction in transmission mode photocathodes. *Applied Physics Letters* **108**, doi:10.1063/1.4944790 (2016).

45    Lee, H. *et al.* A cryogenically cooled high voltage DC photoemission electron source. *Review of Scientific Instruments* **89**, doi:10.1063/1.5024954 (2018).

46    Pierce, C. M. *et al.* Experimental Characterization of Photoemission from Plasmonic Nanogroove Arrays. *Physical Review Applied* **19**, 034034, doi:10.1103/PhysRevApplied.19.034034 (2023).

47    Maxson, J. *et al.* Design, conditioning, and performance of a high voltage, high brightness dc photoelectron gun with variable gap. *Review of Scientific Instruments* **85**, doi:10.1063/1.4895641 (2014).

48    Hernandez-Garcia, C. *et al.* Compact -300 kV dc inverted insulator photogun with biased anode and alkali-antimonide photocathode. *Physical Review Accelerators and Beams* **22**, 113401, doi:10.1103/PhysRevAccelBeams.22.113401 (2019).

49    Konomi, T. *et al.* Development of srf gun applying new cathode idea using a transparent superconducting layer. *59th ICFA Advanced Beam Dynamics Workshop on Energy Recovery Linacs*, 1-3, doi:doi:10.18429/JACoW-ERL2017-MOIACC002 (2017).

50    Kostin, R., Jing, C., Posen, S., Khabiboulline, T. & Bice, D. Nb3Sn SRF Photogun High Power Test at Cryogenic Temperatures. *Journal of latex class files* **14**, 1-5 (2021).

51    Posen, S. *et al.* Advances in Nb3Sn superconducting radiofrequency cavities towards first practical accelerator applications. *Supercond. Sci. Technol.* **34**, 025007, doi:10.1088/1361-6668/abc7f7 (2021).

52    Kostin, R., Jing, C., Posen, S., Khabiboulline, T. & Bice, D. in *5th North American Particle Accel. Conf.* 607-610 (JACoW Publishing).

53    Vicario, C. *et al.* in *Proceedings of FEL2010.* 425-428.

54    Bartnik, A., Gulliford, C., Hoffstaetter, G. H. & Maxson, J. Ultimate bunch length and emittance performance of an MeV ultrafast electron diffraction apparatus with a dc gun and a multicavity superconducting rf linac. *Physical Review Accelerators and Beams* **25**, 093401, doi:10.1103/PhysRevAccelBeams.25.093401 (2022).

55    Pulsar. *Pulsar Physics and the General Particle Tracer (GPT) code*, <https://www.pulsar.nl/gpt/> (2022).

56	van der Geer, S. B., Luiten, O. J., de Loos, M. J., Poplau, G. & van Rienen, U. 3D space-charge model for GPT simulations of high-brightness electron bunches. *Inst. Phys. Conf. Ser.* **175**, 101-110 (2005).
57	Zuo, J.M., and Spence, J.C.H., Advanced transmission electron microscopy, Springer, New York, 2017, ISBN 978-1-4939-6605-9.
58	Photonics, N. *taccor – Ultrafast Femtosecond Lasers*, <https://novantaphotonics.com/product/taccor-ultrafast-femtosecond-lasers/> (2022).

# APPENDICES

|            | L (mm) | Gradient (SRF, T/m) | Gradient (DC, T/m) |
|------------|--------|---------------------|--------------------|
| Quadrupole | 2.0    | -504.1              | -562.7             |
| Drift      | 0.8    |                     |                    |
| Quadrupole | 3.0    | 981.3               | 1095.4             |
| Drift      | 0.8    |                     |                    |
| Quadrupole | 3.8    | -986.2              | -1100.9            |
| Drift      | 0.8    |                     |                    |
| Quadrupole | 3.0    | 981.3               | 1095.4             |
| Drift      | 0.8    |                     |                    |
| Quadrupole | 2.0    | -504.1              | -562.7             |

Table S1. Layout and gradients of the objective lens. The pole tip radius is 0.5 mm. The focal length is 1.00 cm for both the cases of the SRF and the DC guns.

|            | L (mm) | Gradient (SRF, T/m) | Gradient (DC, T/m) |
|------------|--------|---------------------|--------------------|
| Quadrupole | 6      | -36.2 – -30.6       | -55.8 – -34.2      |
| Drift      | 5      |                     |                    |
| Quadrupole | 14     | 43.0 – 37.3         | 61.8 – 41.7        |
| Drift      | 5      |                     |                    |
| Quadrupole | 16     | -46.6 – -41.3       | -63.5 – -46.1      |
| Drift      | 5      |                     |                    |
| Quadrupole | 14     | 43.0 – 37.3         | 61.8 – 41.7        |
| Drift      | 5      |                     |                    |
| Quadrupole | 6      | -36.2 – -30.6       | -55.8 – -34.2      |

Table S2. Layout and gradients of the condenser lenses 1 and 2. The pole tip radius is 1 mm.

|            | L (mm) | Gradient (SRF, T/m) | Gradient (DC, T/m) |
|------------|--------|---------------------|--------------------|
| Quadrupole | 6      | -16.9 – -30.7       | -18.7 – -34.3      |
| Drift      | 5      |                     |                    |
| Quadrupole | 14     | 21.6 – 37.4         | 23.9 – 41.7        |
| Drift      | 5      |                     |                    |
| Quadrupole | 16     | -24.9 – -41.3       | -27.6 – -46.1      |
| Drift      | 5      |                     |                    |
| Quadrupole | 14     | 21.6 – 37.4         | 23.9 – 41.7        |
| Drift      | 5      |                     |                    |
| Quadrupole | 6      | -16.9 – -30.7       | -18.7 – -34.3      |

Table S3. Layout and gradients of the condenser lenses 3. The pole tip radius is 1 mm.